\documentstyle[amstex,12pt]{article}
\begin{document}

\author{{\bf Fabio Cardone}$^{a,b}${\bf , Alessio Marrani}$^{c,d}${\bf \ } \and {\bf %
and Roberto Mignani}$^{b-d}$ \\
$a$ Dipartimento di Fisica\\
Universit\`{a} dell'Aquila\\
Via Vetoio \\
67010 COPPITO, L'Aquila, Italy\\
$b$ I.N.D.A.M. - G.N.F.M.\\
$c$ Universit\`{a} degli Studi ''Roma Tre''\\
Via della Vasca Navale, 84\\
I-00146 ROMA, Italy\\
$d$ I.N.F.N. - Sezione di Roma III}
\title{{\bf The electron mass from Deformed Special Relativity}}
\maketitle

\begin{abstract}
Deformed Special Relativity (DSR) is a generalization of Special Relativity
based on a deformed Minkowski space, i.e. a four-dimensional space-time with
metric coefficients depending on the energy. We show that, in the DSR
framework, it is possible to derive the value of the electron mass from the
space-time geometry via the experimental knowledge of the parameter of local
Lorentz invariance breakdown, and of the Minkowskian threshold energy $%
E_{0,em}$ for the electromagnetic interaction.
\end{abstract}

\bigskip \pagebreak

\section{Introduction\protect\bigskip}

In the last years, two of the present authors (F.C. and R.M.) proposed a
generalization\medskip\ of {\em Special Relativity} (SR) based on a\medskip\
''deformation'' of space-time, assumed to be endowed with a metric whose
coefficients depend on the energy of\medskip\ the process considered [1].
Such a formalism ({\em Deformed Special Relativity}, DSR) applies in
principle to {\em all\medskip } four interactions (electromagnetic, weak,
strong and gravitational) - at least as far as their nonlocal behavior
and\medskip\ nonpotential part is concerned - and provides a metric
representation of them (at\medskip\ least for the process and in
the\medskip\ energy range considered) ([1]-[4], [7], [21] and [24]-[26]).
Moreover, it was shown that such a formalism is actually a\medskip\
five-dimensional one, in the\medskip\ sense that the deformed Minkowski
space is embedded in\medskip\ a larger Riemannian manifold, with energy as
fifth dimension [5].\medskip

In this paper, we will show that the DSR formalism yields an expression of
the electron mass $m_{e}$ in terms of the parameter $\delta $ of local
Lorentz invariance (LLI) breakdown and of the threshold energy for the
gravitational metric, $E_{0,grav}$ (i.e. the energy value under which the
metric becomes Minkowskian). This allows us to evaluate $m_{e}$ from the
(experimental) knowledge of such parameters.

The organization of the paper is as follows. In Sect. 2 we briefly introduce
the concept of deformed Minkowski space, and give the explicit forms of the
phenomenological energy-dependent metrics for the four fundamental
interactions. The LLI breaking parameter $\delta _{int}$ for a given
interaction is introduced in Sect. 3. In Sect. 4 we assume the existence of
a stable fundamental particle interacting gravitationally,
electromagnetically and weakly, and show (by imposing some physical
requirements) that its mass value (expressed in terms of $\delta _{e.m.}$
and $E_{0,grav}$) is just the electron mass. Sect. 5 concludes the
paper.\pagebreak

\section{\protect\bigskip Deformed Special Relativity in four dimensions
(DSR4)}

\subsection{Deformed Minkowski space-time}

The generalized (``deformed'') Minkowski space $\widetilde{M_{4}}$ (DMS4) is
defined as a space with the same local coordinates $x$ of $M_{4}$ (the
four-vectors of the usual Minkowski space), but with metric given by the
metric tensor\footnote{%
In the following, we shall employ the notation ''ESC on'' (''ESC off'') to
mean that the Einstein sum convention on repeated indices is (is not) used.}
\begin{eqnarray}
\eta _{\mu \nu }(E) &=&diag\left(
b_{0}^{2}(E),-b_{1}^{2}(E),-b_{2}^{2}(E),-b_{3}^{2}(E)\right) =\smallskip
\nonumber \\
&&  \nonumber \\
&&\stackrel{\text{{\footnotesize ESC off}}}{=}\delta _{\mu \nu }\left[
\delta _{\mu 0}b_{0}^{2}(E)-\delta _{\mu 1}b_{1}^{2}(E)-\delta _{\mu
2}b_{2}^{2}(E)-\delta _{\mu 3}b_{3}^{2}(E)\right] \text{ }
\end{eqnarray}
($\forall E\in R_{0}^{+}$), where the $\left\{ b_{\mu }^{2}(E)\right\} $ are
dimensionless, real, positive functions of the energy [1]. The generalized
interval in $\widetilde{M_{4}}$ is therefore given by ($x^{\mu
}=(x^{0},x^{1},x^{2},x^{3})=(ct,x,y,z)$, with $c$ being the usual light
speed in vacuum) (ESC on)
\begin{equation}
ds^{2}=b_{0}^{2}(E)c^{2}dt^{2}-(b_{1}^{2}(E)dx^{2}+b_{2}^{2}(E)dy^{2}+b_{3}^{2}(E)dz^{2})=\eta _{\mu \nu }(E)dx^{\mu }dx^{\nu }=dx\ast dx.
\end{equation}

The last step in (2) defines the scalar product $\ast $ in the deformed
Minkowski space $\widetilde{M_{4}}$ \footnote{%
Notice that our formalism - in spite of the use of the word ''deformation''
- has nothing to do with the ''deformation'' of the Poincar\'{e} algebra
introduced in the framework of quantum group theory (in particular the
so-called $\kappa $-deformations) [6]. In fact, the quantum group
deformation is essentially a modification of the commutation relations of
the Poincar\'{e} generators, whereas in the DSR framework the deformation
concerns the metrical structure of the space-time (although the Poincar\'{e}
algebra is affected, too [7]).}.\bigskip\ It follows immediately that it can
be regarded as a particular case of a Riemann space with null curvature.

Let us stress that, in this formalism, the energy $E$ is to be understood as
the {\em energy of a physical process} measured by the detectors via their
electromagnetic interaction in the usual Minkowski space. Moreover, $E$ is
to be considered as a {\em dynamical variable} (on the same footing as the
space-time coordinates), because it specifies the {\em dynamical behavior}
of the process under consideration, and, via the metric coefficients, it
provides us with a {\em dynamical map} - in the energy range of interest -
of the interaction ruling the given process. Let's recall that the use of
{\em momentum components as dynamical variables} on the same foot of the
space-time ones can be traced back to Ingraham [8]. Dirac [9], Hoyle and
Narlikar [10] and Canuto et al. [11] treated mass as a dynamical variable in
the context of scale-invariant theories of gravity.

Moreover - as already stressed in the Introduction - the 4-d. deformed
Minkowski space can be {\em naturally embedded} in a 5-d. Riemann space,
with energy as fifth metrical coordinate [5]. Curved 5-d. spaces have been
considered by several Authors [12]. On this respect, the DSR formalism is a
kind of generalized ({\em non-compactified}) Kaluza-Klein theory, and
resembles, in some aspects, the so-called ''Space-Time-Mass'' (STM) theory
(in which the fifth dimension is the rest mass), proposed by Wesson [13] and
studied in detail by a number of Authors [14].

\subsection{Energy-dependent phenomenological metrics for the four
interactions}

As far as the phenomenology is concerned, we recall that a {\em local}
breakdown of Lorentz invariance may be envisaged for all four fundamental
interactions (electromagnetic, weak, strong and gravitational) whereby {\em %
one\thinspace\ gets \thinspace evidence \thinspace for \thinspace
a\thinspace\ departure of \thinspace the\thinspace\ space-time \thinspace
metric \thinspace from \thinspace the \thinspace Minkowskian\thinspace
one\thinspace } (at least in the energy range examined). The experimental
data analyzed were those of the following four physical processes:

- the lifetime of the (weakly decaying) $K_{s}^{0}$ meson [15];

- the Bose-Einstein correlation in (strong) pion production [16];

- the superluminal photon tunneling [17];

- the comparison of clock rates in the gravitational field of Earth [18].

A detailed derivation and discussion of the energy-dependent
phenomenological metrics for all the four interactions can be found in Ref.s
[1]-[4]. Here, we confine ourselves to recall their following basic features:

{\bf 1)} Both the {\bf electromagnetic} and the {\bf weak} metric show the
same functional behavior, namely

\begin{equation}
\eta (E)=diag(1,-b^{2}(E),-b^{2}(E),-b^{2}(E));  \label{emweak1}
\end{equation}

\begin{eqnarray}
b^{2}(E) &=&\left\{
\begin{array}{lll}
(E/E_{0})^{1/3}, & 0<E\leq E_{0} &  \\
&  &  \\
1, & E_{0}<E &
\end{array}
\right. = \\
&&  \nonumber \\
&&  \nonumber \\
&=&1+\theta (E_{0}-E)\left[ \left( \frac{E}{E_{0}}\right) ^{1/3}-1\right]
,E>0,  \label{emweak2}
\end{eqnarray}
(where $\theta (x)$ is the Heavyside theta function) with the only
difference between them being the threshold energy $E_{0}$, i.e. the energy
value at which the metric parameters are constant, i.e. the metric becomes
Minkowskian ($\eta _{\mu \nu }(E\geq E_{0})\equiv g_{\mu \nu
}=diag(1,-1,-1,-1)$); the fits to the experimental data yield

\begin{gather}
E_{0,e.m.}=\left( 4.5\pm 0.2\right) \mu eV\,;  \nonumber \\
\nonumber \\
E_{0,weak}=\left( 80.4\pm 0.2\right) GeV.
\end{gather}
Notice that for either interaction the metric is isochronous, spatially
isotropic and {\em ''sub-Minkowskian''}, i.e. it approaches the Minkowskian
limit from below (for $E<E_{0}$). Both metrics are therefore Minkowskian for
$E>E_{0,weak}\simeq 80GeV$, and then our formalism is fully consistent with
electroweak unification, which occurs at an energy scale $\sim 100GeV$.

Let us recall that the phenomenological electromagnetic metric (3)-(5) was
derived by analyzing the propagation of evanescent waves in undersized
waveguides [16]. \ It allows one to account for the observed superluminal
group speed in terms of a nonlocal behavior of the waveguide, just described
by an effective deformation of space-time in its reduced part [3]. As to the
weak metric, it was obtained by fitting the data on the meanlife of the
meson $K_{s}^{0}$ (experimentally known in a wide energy range $(30\div
350GeV)$ [14]), thus accounting for its apparent departure from a purely
Lorentzian behavior ([1], [19]).

{\bf 2)} For the {\bf strong} interaction, the metric was derived [2] by
analyzing the phenomenon of Bose-Einstein (BE) correlation for $\pi $-mesons
produced in high-energy hadronic collisions [16]. Such an approach permits
to describe the BE effect as the decay of a ''fireball'' whose lifetime and
space sizes are directly related to the metric coefficients $\left\{ b_{\mu
,strong}^{2}(E)\right\} $, and to avoid the introduction of ''ad hoc''
parameters in the pion correlation function [2]. The strong metric reads
\begin{equation}
\eta
_{strong}(E)=diag(b_{0,strong}^{2}(E),-b_{1,strong}^{2}(E),-b_{2,strong}^{2}(E),-b_{3,strong}^{2}(E));
\end{equation}
\begin{eqnarray}
b_{1,strong}^{2}(E) &=&\left( \frac{\sqrt{2}}{5}\right) ^{2};  \nonumber \\
&&  \nonumber \\
b_{2,strong}^{2}(E) &=&\left( \frac{2}{5}\right) ^{2},\forall E>0;
\end{eqnarray}

\begin{eqnarray}
b_{0,strong}^{2}(E) &=&b_{3,strong}^{2}(E)=\left\{
\begin{array}{lll}
1, & 0<E\leq E_{0,strong} &  \\
&  &  \\
(E/E_{0,strong})^{2}, & E_{0,strong}<E &
\end{array}
\right. =  \nonumber \\
&& \\
&&  \nonumber \\
&=&\smallskip 1+\theta (E-E_{0,strong})\left[ \left( \frac{E}{E_{0,strong}}%
\right) ^{2}-1\right] ,E>0
\end{eqnarray}
with

\begin{equation}
E_{0,strong}=\left( 367.5\pm 0.4\right) GeV.
\end{equation}
\noindent Let us stress that, in this case, contrarily to the
electromagnetic and the weak ones, {\em a \thinspace deformation \thinspace
of \thinspace the \thinspace time \thinspace coordinate\thinspace occurs};
moreover, {\em the\thinspace three-space is\thinspace\ anisotropic}{\it ,}
with two spatial parameters constant (but different in value) and the third
one variable with energy like the time one.

{\bf 3)} The {\bf gravitational} energy-dependent metric was obtained [4] by
fitting the experimental data on the relative rates of clocks in the Earth
gravitational field [18]. Its explicit form is\footnote{%
The coefficients $b_{1,grav.}^{2}(E)$\ and $b_{2,grav.}^{2}(E)$\ are
presently {\em undetermined} at phenomenological level.}:

\begin{equation}
\eta
_{grav.}(E)=diag(b_{0,grav.}^{2}(E),-b_{1,grav.}^{2}(E),-b_{2,grav.}^{2}(E),-b_{3,grav.}^{2}(E));
\label{grav1}
\end{equation}

\begin{eqnarray}
b_{0,grav.}^{2}(E) &=&b_{3,grav.}^{2}(E)=\left\{
\begin{array}{lll}
1, &  & 0<E\leq E_{0,grav.} \\
&  &  \\
\frac{1}{4}(1+E/E_{0,grav.})^{2}, &  & E_{0,grav.}<E
\end{array}
\right. =  \nonumber \\
&&  \nonumber \\
&&  \nonumber \\
&=&1+\theta (E-E_{0,grav.})\left[ \frac{1}{4}\left( 1+\frac{E}{E_{0,grav.}}%
\right) ^{2}-1\right] ,E>0
\end{eqnarray}
with
\begin{equation}
E_{0,grav.}=\left( 20.2\pm 0.1\right) \mu eV.  \label{grav2}
\end{equation}
Intriguingly enough, this is approximately of the same order of magnitude of
the thermal energy corresponding to the $2.7^{o}K$ cosmic background
radiation in the Universe\footnote{%
It is worth stressing that the energy-dependent gravitational metric
(10)-(12) is to be regarded as a {\em local} representation of gravitation,
because the experiments considered took place in a neighborhood of Earth,
and therefore at a small scale with respect to the usual ranges of gravity
(although a large one with respect to the human scale).}.

Notice that the strong and the gravitational metrics are {\em %
over-Minkowskian} (namely, they approach the Minkowskian limit from above ($%
E_{0}<E$), at least for their coefficients $b_{0}^{2}(E)=b_{3}^{2}(E)$).

\section{LLI breaking factor and relativistic energy in DSR}

The breakdown of standard local Lorentz invariance (LLI) is expressed by the
LLI breaking factor parameter $\delta $ [19]. We recall that two different
kinds of LLI violation parameters exist: the {\em isotropic} (essentially
obtained by means of experiments based on the propagation of e.m. waves,
e.g. of the Michelson-Morley type), and the {\em anisotropic} ones (obtained
by experiments of the Hughes-Drever type [19], which test the isotropy of
the nuclear levels).

In the former case, the LLI violation parameter reads [19]
\begin{eqnarray}
\delta &=&\left( \frac{u}{c}\right) ^{2}-1, \\
u &=&c+v,  \nonumber
\end{eqnarray}
where $c$ is, as usual, the speed of light {\em in vacuo}, $v$ is the LLI
breakdown speed (e.g. the speed of the preferred frame) and $u$ is the new
speed of light (i.e. the {\em ''maximal causal speed''} in Deformed Special
Relativity [1]). In the {\em anisotropic} case, there are different
contributions $\delta ^{A}$\ to the anisotropy parameter from the different
interactions. In the HD experiment, it is $A=S,HF,ES,W$, meaning strong,
hyperfine, electrostatic and weak, respectively. These correspond to four
parameters $\delta ^{S}$ (due to the strong interaction), $\delta ^{ES}$
(related to the nuclear electrostatic energy), $\delta ^{HF}$ (coming from
the hyperfine interaction between the nuclear spins and the applied external
magnetic field) and $\delta ^{W}$ (the weak interaction contribution).

In our framework, we can define $\delta $ as follows:
\begin{equation}
\delta _{int.}\equiv \frac{m_{in.,int.}-m_{in.,grav.}}{m_{in.,int.}}=1-\frac{%
m_{in.,grav.}}{m_{in.,int.}},  \label{delta}
\end{equation}
where $m_{in.,int.}$ is the inertial mass of the particle considered with
respect to the given interaction \footnote{%
Throughout the present work, $"int."$ denotes a physically detectable
fundamental interaction, which can be operationally defined by means a
phenomenological energy-dependent metric of deformed Minkowskian type.}. In
other words, we assume that the {\em local} deformation of space-time
corresponding to the interaction considered, and described by the metric
(1), gives rise to a {\em local violation of the Principle of Equivalence}
for interactions different from the gravitational one. Such a departure,
just expressed by the parameter $\delta _{int.}$, does constitute also a
measure of the amount of LLI breakdown. In the framework of DSR, $\delta
_{int.}$ embodies the {\em geometrical contribution to the inertial mass},
thus discriminating between two different metric structures of space-time.

Of course, if the interaction considered is the gravitational one, the
Principle of Equivalence strictly holds, i.e.
\begin{equation}
m_{in.,grav.}=m_{g},  \label{EP}
\end{equation}
where $m_{g}$ is the gravitational mass of the physical object considered,
i.e. it is its {\em ''gravitational charge''} (namely, its coupling constant
to the gravitational field).

Then, we can rewrite (\ref{delta}) as:
\begin{equation}
\delta _{int.}\equiv \frac{m_{in.,int.}-m_{g}}{m_{in.,int.}}=1-\frac{m_{g}}{%
m_{in.,int.}},  \label{delta-int}
\end{equation}
and therefore, when the particle is subjected {\em only} to gravitational
interaction, it is

\begin{equation}
\delta _{grav.}=0
\end{equation}

\bigskip

In DSR the relativistic energy, for a particle subjected to a given
interaction and moving along $\widehat{x^{i}}$ , has the form [1]:
\begin{eqnarray}
E_{int.} &=&m_{in.,int.}u_{i,int.}^{2}(E)\widetilde{\gamma }_{int.}(E)=
\nonumber \\
&&  \nonumber \\
&=&m_{in.,int.}c^{2}\frac{b_{0,int.}^{2}(E)}{b_{i,int.}^{2}(E)}\left[
1-\left( \frac{v_{i}b_{i,int.}(E)}{cb_{0,int.}(E)}\right) ^{2}\right]
^{-1/2},  \label{En1}
\end{eqnarray}
where $\underline{u}_{int.}(E)$ is the {\em maximal causal velocity} for the
interaction considered (i.e. the analogous of the light speed in SR), given
by ([1],[21])
\begin{equation}
\underline{u}_{int.}(E)\equiv \left( c\frac{b_{0,int.}(E)}{b_{1,int.}(E)},c%
\frac{b_{0,int.}(E)}{b_{2,int.}(E)},c\frac{b_{0,int.}(E)}{b_{3,int.}(E)}%
\right) .  \label{u}
\end{equation}
In the non-relativistic (NR) limit of DSR, i.e. at energies such that
\begin{equation}
v_{i}\ll u_{i,int.}(E),
\end{equation}
Eq. (\ref{u}) yields the following NR expression of the energy corresponding
to the given interaction:
\begin{equation}
E_{int.,NR}=m_{in.,int.}u_{i,int.}^{2}(E)=m_{in.,int.}c^{2}\frac{%
b_{0,int.}^{2}(E)}{b_{i,int.}^{2}(E)}.  \label{EnNR}
\end{equation}

In the case of the gravitational metric (\ref{grav1})-(\ref{grav2}), we have
\begin{equation}
\frac{b_{0,grav.}^{2}(E)}{b_{3,grav.}^{2}(E)}=1,\forall E\in R_{0}^{+}.
\end{equation}
Therefore, for $i=3$ , Eq.s (\ref{En1}) and (\ref{EnNR}) become,
respectively ($v_{3}=v$):
\begin{equation}
E_{grav.}=m_{g}c^{2}\left[ 1-\left( \frac{v}{c}\right) ^{2}\right]
^{-1/2}=m_{g}c^{2}\gamma ,
\end{equation}
\begin{equation}
E_{grav.,NR}=m_{g}c^{2},
\end{equation}
namely, the gravitational energy takes its {\em standard,
special-relativistic }values.

This means that the special characterization (corresponding to the choice $%
i=3$) of Eq.s (\ref{En1}) and (\ref{EnNR}) within the framework of DSR
relates the gravitational interaction with SR, which is - as well known -
based on the electromagnetic interaction in its Minkowskian form.

\section{The electron as a fundamental particle and its ''geometrical'' mass}

Let us now consider for $E$ the threshold energy of the gravitational
interaction:

\begin{equation}
E=E_{0,grav.}
\end{equation}
where $E_{0,grav.}$ is the limit value under which the metric $\eta _{\mu
\nu ,grav.}(E)$ becomes Minkowskian (at least in its known components).
Indeed, from Eq.s (\ref{grav1})-(\ref{grav2}) it follows (\ $\forall E\in
(0,E_{0,grav.}]$):
\begin{eqnarray*}
\eta _{\mu \nu ,grav.}(E)
&=&diag(1,-b_{1,grav.}^{2}(E),-b_{2,grav.}^{2}(E),-1)\stackrel{\text{%
{\footnotesize ESC off}}}{=} \\
&&\smallskip \\
&=&\delta _{\mu \nu }\left[ \delta _{\mu 0}-\delta _{\mu
1}b_{1,grav.}^{2}(E)-\delta _{\mu 2}b_{2,grav.}^{2}(E)-\delta _{\mu 3}\right]
.
\end{eqnarray*}
Notice that at the energy $E=E_{0,grav.}$ the electromagnetic metric (\ref
{emweak1})-(\ref{emweak2}) is Minkowskian, too (because $%
E_{0,grav.}>E_{0,e.m.}$).

On the basis of the previous considerations, it seems reasonable to assume
that the physical object (particle) $p$ with a rest energy (i.e.
gravitational mass) just equal to the threshold energy $E_{0,grav.}$, namely
\begin{equation}
E_{0,grav.}=m_{g,p}c^{2},  \label{E0grav}
\end{equation}
must play a fundamental role for either e.m. and gravitational interaction.
We can e.g. hypothesize that $p$ corresponds to the lightest mass eigenstate
which experiences both force fields (i.e., from a quantum viewpoint,
coupling to the respective interaction carriers, the photon and the
graviton). As a consequence, $p$ must be {\em intrinsically stable}, due to
the impossibility of its decay in lighter mass eigenstates, even in the case
such a particle is subject to weak interaction, too (i.e. it couples to all
gauge bosons of the Glashow-Weinberg-Salam group $SU(2)\otimes U(1)$, not
only to its electromagnetic charge sector\footnote{%
For precision's sake, it should be noticed that actually the physically
consistently-acting gauge group of the (unbroken) Glashow-Weinberg-Salam
electroweak theory is not $SU(2)_{T}\otimes U(1)_{Y}$, but rather
\[
\left( SU(2)\otimes U(1)\right) /Z_{2}\approx U(2),
\]
where $T$ and $Y$ respectively stand for weak isospin and hypercharge
symmetries, and $\otimes $ is the usual direct group product. This cosetting
by the discrete symmetry $Z_{2}$ is due to the very field content of the
actual electroweak theory, as rigorously explained in [23].}).

Since, as we have seen, for $E=E_{0,grav.}$ the electromagnetic metric is
Minkowskian, too, it is natural to assume, for $p$:
\begin{equation}
m_{in.,p,e.m.}=m_{in.,p}
\end{equation}
namely {\em its inertial mass is that measured with respect to the
electromagnetic metric}.

Then, due to the Equivalence Principle (see Eq. (\ref{EP})), the mass of $p$
is characterized by
\begin{equation}
p:\left\{
\begin{array}{c}
m_{in.,p,grav.}=m_{g,p} \\
\\
m_{in.,p,e.m.}=m_{in.,p}.
\end{array}
\right.
\end{equation}

Therefore, for such a fundamental particle the SSLI breaking factor (\ref
{delta-int}) of the e.m. interaction becomes:
\begin{equation}
\delta _{e.m.}=\frac{m_{in.,p}-m_{g,p}}{m_{in.,p}}=1-\frac{m_{g,p}}{m_{in.,p}%
}\Leftrightarrow m_{g,p}=m_{in.,p}\left( 1-\delta _{e.m.}\right) .
\label{mgp}
\end{equation}
Replacing (\ref{mgp}) in (\ref{E0grav}) yields:
\begin{eqnarray}
E_{0,grav.} &=&m_{in.,p}\left( 1-\delta _{e.m.}\right) c^{2}\Leftrightarrow
\nonumber \\
&&  \nonumber \\
&\Leftrightarrow &m_{in.,p}=\frac{E_{0,grav.}}{c^{2}}\frac{1}{1-\delta
_{e.m.}}.  \label{final1}
\end{eqnarray}
Thus, the obtained result allows us to evaluate the inertial mass of $p$
from the knowledge of the electromagnetic LLI breaking parameter $\delta
_{e.m.}$ and of the threshold energy $E_{0,grav.}$ of the gravitational
metric.

The lowest limit to the LLI breaking factor of electromagnetic interaction
has been recently determined by an experiment based on the detection of a DC
voltage across a conductor induced by the steady magnetic field of a coil
[22]. The value found in [22] corresponds to
\begin{equation}
1-\delta _{e.m.}\widetilde{=}4\cdot 10^{-11}.  \label{DC}
\end{equation}
Then, inserting the value (\ref{grav2}) for $E_{0,grav.}$ \footnote{%
Let us recall that the value of $E_{0,grav}.$ was determined by fitting the
experimental data on the slowing down of clocks in the Earth gravitational
field [18]. See also Ref. [4].}and (\ref{DC}) in (\ref{final1}), we get
\begin{equation}
m_{in.,p}=\frac{E_{0,grav.}}{c^{2}}\frac{1}{1-\delta _{e.m.}}\geq \frac{%
2\cdot 10^{-5}}{4\cdot 10^{-11}}\frac{eV}{c^{2}}=0.5\frac{MeV}{c^{2}}%
=m_{in.,e}  \label{final2}
\end{equation}
(with $m_{in,e}$ being the inertial electron mass), where the $\geq $ is due
to the fact that in general the LLI breaking factor constitutes an {\em %
upper limit } (i.e. it sets the scale {\em under which }a violation of LLI
is expected).

If experiment [22] {\em does indeed provide evidence }for a LLI breakdown
(as it seems the case, although further confirmation is needed), Eq. (\ref
{final2}) yields
\begin{equation}
m_{in.,p}=m_{in.,e}.
\end{equation}
We find therefore the amazing result that {\em the fundamental particle }$p$%
{\em \ is nothing but the electron }$e^{-}${\em \ (or its antiparticle }$%
e^{+}$ \footnote{%
Of course, this last statement does strictly holds only if the CPT Theorem
mantains its validity in the DSR framework, too. Although this problem has
not yet been addressed in general on a formal basis, we can state that it
holds true in the case we considered, since we assumed that the energy value
is $E=E_{0,grav.}$, corresponding to the Minkowskian form of both
electromagnetic and gravitational metric.}{\em ). }The electron is indeed
the lightest massive lepton (pointlike, non-composite particle) with
electric charge, and therefore subjected to gravitational, electromagnetic
and weak interactions, but unable to weakly decay due to its small mass.
Consequently, $e^{-}$ ($e^{+}$) shares all the properties we required for
the particle $p$, whereby it plays a fundamental role for gravitational and
electromagnetic interactions.

\bigskip

\section{Conclusions}

The formalism of DSR describes -among the others -, in geometrical terms
(via the energy-dependent deformation of the Minkowski metric) the breakdown
of Lorentz invariance at {\em local} level (parametrized by the LLI breaking
factor $\delta _{int.}$). We have shown that within DSR it is possible - on
the basis of simple and plausible assumptions - to evaluate the inertial
mass of the electron $e^{-}$ (and therefore of its antiparticle, the
positron $e^{+}$) by exploiting the expression of the relativistic energy in
the deformed Minkowski space $\widetilde{M_{4}}(E)_{E\in R_{0}^{+}}$ , the
explicit form of the phenomenological metric describing the gravitational
interaction (in particular its threshold energy), and the LLI breaking
parameter for the electromagnetic interaction $\delta _{e.m.}$ .

Therefore, {\em the inertial properties of one of the fundamental
constituents of matter and of Universe do find a ''geometrical''
interpretation in the context of DSR, by admitting for local violations of
standard Lorentz invariance}.\pagebreak

\end{document}